%% file: 00_main.tex
\documentclass {llncs}

\usepackage{booktabs} 

\usepackage{pifont}
\usepackage[english]{babel}
\usepackage[utf8x]{inputenc}
\usepackage{amsmath}
\usepackage{graphicx}
\usepackage{todonotes}
\usepackage{url}
\usepackage{subcaption}
\usepackage{diagbox}
\usepackage{soul} 
\usepackage{bm}
\usepackage{footnote}
\usepackage{enumitem}

\usepackage{floatrow}

\usepackage[export]{adjustbox}

\definecolor{lightgreen}{rgb}{0.6, 1, 0.6}
\definecolor{lightred}{rgb}{1, 0.6, 0.6}
\DeclareRobustCommand\parisnew[1]{#1}

\newcommand\paris[0]{$\mathcal{P}$}
\newcommand\brexit[0]{$\mathcal{B}$}
\newcommand\brexitbold[0]{$\bm{\mathcal{B}}$}
\newcommand\parisbold[0]{$\bm{\mathcal{P}}$}

\newcommand\bwd[0]{$\mathcal{B}.wd$}
\newcommand\bwdbold[0]{$\mathcal{B}.wd$}

\newcommand\bref[0]{$\mathcal{B}.ref$}
\newcommand\brefbold[0]{$\mathcal{B}.ref$}

\newcommand\pa[0]{$\mathcal{P}.agr$}
\newcommand\pabold[0]{$\mathcal{P}.agr$}

\newcommand\pwd[0]{$\mathcal{P}.wd$}
\newcommand\pwdbold[0]{$\mathcal{P}.wd$}

\newcommand{\no}{}

\begin{document}

\title{
Towards better Understanding Researcher Strategies in  
Cross-lingual Event Analytics}

\author{Simon Gottschalk\inst{1} \and Viola Bernacchi\inst{2} \and Richard Rogers\inst{3} \and Elena Demidova\inst{1}}

\authorrunning{S. Gottschalk, V. Bernacchi, R. Rogers, E. Demidova}

\institute{L3S Research Center, Leibniz Universität Hannover, Hannover, Germany \\
\email{\{gottschalk,demidova\}@L3S.de}
\and DensityDesign Research Lab, Milano, Italy \\
\email{viola.bernacchi@live.it}
\and University of Amsterdam, Amsterdam, The Netherlands \\
\email{rogers@uva.nl}}

\maketitle

\begin{abstract}
With an increasing amount of information on globally important events, there is a growing demand for efficient analytics of multilingual event-centric information. Such analytics is particularly challenging due to the large amount of content, the event dynamics and the language barrier. Although memory institutions increasingly collect event-centric Web content in different languages, very little is known about the strategies of researchers who conduct analytics of such content. 
In this paper we present researchers’ strategies for the content, method and feature selection in the context of cross-lingual event-centric analytics observed in two case studies on multilingual Wikipedia. We discuss the influence factors for these strategies, the findings enabled by the adopted methods along with the current limitations and provide recommendations for services supporting researchers in cross-lingual event-centric analytics.
\end{abstract}

\input{01_introduction}

\input{02_objectives}

\input{03_case_study}

\input{04_results_and_observations}

\input{06_discussion_new}

\input{07_related_work}

\input{08_conclusion}

\bibliographystyle{splncs04}

{
\scriptsize
\subsubsection*{Acknowledgements} This work was partially funded by the ERC ("ALEXANDRIA", 339233) and H2020-MSCA-ITN-2018-812997 "Cleopatra".
\bibliography{references} 
}

\end{document}

%% file: 01_introduction.tex
\section{Introduction}
\label{sec:introduction}

The world's community faces an unprecedented number of events impacting it as a whole across language and country borders. Recently, such unexpected events included political shake-ups such as Brexit and the US pullout of the Paris Agreement. 
Such events result in a vast amount of event-centric, multilingual information that differs across sources, languages and communities and can reflect community-specific aspects such as opinions, sentiments and bias \cite{Rogers:2013}.  
In the context of events with global impact, cross-cultural studies gain in importance.

Memory institutions are increasingly interested in collecting multilingual event-centric information and making this information available to interested researchers. 
For example, the Internet Archive provides the Archive-It service that facilitates curated collections of Web content\footnote{\url{archive-it.org}}. 
Several recent research efforts target the automatic creation of event-centric collections from the Web and large-scale Web archives (e.g. iCrawl \cite{GossenDR15}, \cite{GossenDR17})
as well as creation of event-centric knowledge graphs such as EventKG \cite{GottschalkD18}. 
In this context one of the key Web resources to analyze cross-cultural and cross-lingual differences in representations of current and historical events is the multilingual Wikipedia \cite{Pentzold:2017}, \cite{Rogers:2013}. 

However, at present 
very little is known about the strategies and the requirements of researchers who analyze event-centric cross-lingual information.
In this paper we take the first important step towards a better understanding of researcher strategies in the context of event-centric cross-lingual studies at the example of multilingual Wikipedia. 
The goals of this paper are to better understand: 
1) How do researchers analyze current events in multilingual settings? In particular, we are interested in the content selection strategies, analysis methods and features adopted along with the influence factors for this adoption.
2) Which findings can be facilitated through existing cross-lingual analytics methods, what limitations do these methods have and how to overcome them?

To address these questions we conducted two qualitative case studies that concerned the Brexit referendum and the US pullout of the Paris Agreement.
We observed interdisciplinary and multicultural research teams who performed analyses of the event representations in the multilingual Wikipedia dataset during a week's time. 
As a first step, we used in-depth pre-session questionnaires aimed at collecting the participants' background. Following that, the participants defined their own research questions and several working sessions took place. 
During these sessions we observed the methods adopted by the participants and the findings they obtained. Finally, we interviewed the participants. 

The main findings of our analysis are as follows: 
1) \textit{The content selection strategy} mostly depends on the event characteristics and the collection properties. 
2) \textit{The adoption of analysis methods and features} is most prominently influenced by the researcher backgrounds, the information structure and the analysis tools. 
3) \textit{The features involved} in the adopted analysis methods mostly include metadata (e.g. tables of content), rather than the actual texts. 
4) \textit{The main insights facilitated by the adopted analysis methods} include a variety of findings e.g. with respect to the shared and language-specific aspects, the interlingual dependencies, the event dynamics, as well as the originality and role of language editions. 
5) \textit{The limitations of the adopted methods} mostly concern 
the relatively low content and temporal resolution, as well as the lack of detailed insights into the communities and discussions behind the content. 
6) \textit{The recommendations to overcome these limitations} include 
the development of tools that better support cross-lingual overview, facilitate fact alignment, provide high temporal resolution, as well as community and discussion insights.

%% file: 02_objectives.tex
\section{Study Context and Objectives}
\label{sec:objectives}

We focus on two political events with global impact that constitute important cases for the cross-cultural analysis:

\begin{itemize}[noitemsep,topsep=0pt,partopsep=0pt]
\item \textbf{B}rexit (\brexit{}): On 23 June 2016, a referendum took place to decide on the withdrawal of the United Kingdom (UK) from the European Union (EU). 51.9\% of the voters voted to leave the EU, which lead to the withdrawal process called “Brex\-it”.
\item US \textbf{P}aris Agreement pullout (\paris{}): On June 1, 2017, the US President Donald Trump announced to pull out of the Paris agreement, which was previously signed by 195 countries at the Paris climate conference.
\end{itemize}

To better understand researcher strategies, 
we asked the participants to conduct an analysis of the event, in particular with the focus on the international event perception.
Overall, three main objectives ($O1$-$O3$) are addressed:

\textit{O1 - Content selection. How do researchers select articles, languages and revisions to analyze, given an ongoing event? Which factors influence this selection?} 
Wikipedia articles are generated in a dynamic process where 
each edit of an article results in a new version called \textit{revision}. 
Given the high velocity of discussions in different Wikipedia language editions surrounding an ongoing global event, there is a large amount of potentially relevant information. 
Thus, there is a need to identify the most relevant articles, their revisions and language versions as entry points for the detailed analysis.

\textit{O2 - Method and feature selection. Which methods and features can researchers use efficiently to perform cross-lingual event-centric analytics? Which factors influence this selection?} 
Wikipedia articles describing significant events tend to cover a large number of aspects.
The large number of articles, their revisions and the variety of language 
editions make the analysis particularly difficult. 
The challenge here is to choose analysis methods and features that can provide an overview of cross-lingual and temporal differences across multilingual event representations efficiently. 

\textit{O3 - Findings and limitations. Which findings can be efficiently obtained by researchers when conducting research over multilingual, event-centric articles using particular analysis methods? What are the current limitations and how can they be addressed?} 
The size, dynamics as well as cross-lingual and cross-cultural nature of Wikipedia articles pose challenges on the interpretation of research results, especially in case such an interpretation requires close reading of multilingual content. Our goal here is to better understand which findings can be obtained efficiently, and derive recommendations for future assistance. 

We do not aim at the completeness of the considered strategies, methods, features and interpretation results, but focus on the 
participants' approaches as a starting point to better understand which methods and features appear most efficient from the participant perspective, which factors influence their selection and which findings they can enable in practice.

%% file: 03_case_study.tex
\section{Methodology}
\label{sec:case_study}

The case studies were conducted by performing the following steps: 
\begin{itemize}[noitemsep,topsep=0pt,partopsep=0pt]
\item [1.] Introduction of the event to the participants. 
\item [2.] Individual questionnaires to be filled out by the participants.
\item [3.] Working sessions in teams, observed by the authors.
\item [4.] Individual semi-structured interviews with the participants.
\end{itemize}

\subsection{Pre-Session Questionnaires}
\label{sec:participants}

\begin{table}[t!]
\footnotesize
\centering
\caption{Setup and the participant background. CS:~Computer scientist, ID:~Information designer, S:~Sociologist.}
\label{tab:participants}
\begin{tabular}{l|l|l|}
\cline{2-3}
 & \multicolumn{1}{c|}{\textbf{\brexitbold{}}} & \multicolumn{1}{c|}{\textbf{\parisbold{}}} \\ \hline \hline\multicolumn{3}{|l|}{\textbf{Study setup}} \\ \hline
\multicolumn{1}{|l|}{\textbf{Event date}} & June 23, 2016 & June 1, 2017 \\ \hline
\multicolumn{1}{|l|}{\textbf{Study dates}} & June 27 - July 1, 2016 & July 3 - July 7, 2017 \\ \hline
\multicolumn{1}{|l|}{\textbf{Overall study duration}} & 14 hours & 14 hours \\ \hline \hline
\multicolumn{3}{|l|}{\textbf{Participant background}} \\ \hline
\multicolumn{1}{|l|}{\textbf{Number of participants}} & 5 (ID: 3, S: 2) & 4 (CS: 1, ID: 2, S: 1) \\ \hline
\multicolumn{1}{|l|}{\textbf{Native languages}} & IT (3), NL (1), UK (1) & IT(2), DE (2) \\ \hline
\multicolumn{1}{|l|}{\textbf{Languages spoken}} & EN, IT, NL, DE & EN, IT, DE, FR, ES \\ \hline \hline
\multicolumn{3}{|l|}{\textbf{Wikipedia experience}} \\ \hline
\multicolumn{1}{|l|}{\textbf{Role}} & reader (5) & reader (4), editor (2) \\ \hline
\multicolumn{1}{|l|}{\textbf{Frequency of use}} & daily (1), weekly (4) & daily (3), weekly (1) \\ \hline
\multicolumn{1}{|l|}{\textbf{Multilingual Wikipedia experience}} & yes (2) & yes (4) \\ \hline
\end{tabular}
\end{table}

Table \ref{tab:participants} provides an overview of the study setup 
and the participant interdisciplinary and multicultural background, collected using pre-session questionnaires. 
According to these questionnaires, the participants estimated the language barrier to be a major problem in both of the studies
and raised the question whether it is possible not only to detect cultural differences or commonalities, but also to reason about their origins.

\subsection{Task Definition and Working Sessions}
\label{sec:task}

We asked the participants to: 1) define their own research questions and analysis methods;
2) conduct an analysis of the event-related articles across
different Wikipedia language editions with respect to these questions; and 
3) present their findings. 
This way, we kept the task description of the study rather open, as we intended to facilitate an open-minded discussion among the participants, 
to enhance their motivation and to reduce possible bias.

To enable in-depth insights, the studies implied high expenditures of approximately 14 hours per participant, which overall translates into 126 person hours.
The participants worked together as a team over four days.

%% file: 04_results_and_observations.tex
\section{The Participant Approach}
\label{sec:approach}

The interdisciplinary expertise of the participants enabled them to tackle several facets of interest in the context of the considered events. 
In this section, we describe and compare the participant course of action in both case studies, from the definition 
of the research questions to the presentation of results.

\subsection{Research Questions}
\label{sec:questions and methods}

At the beginning of the case studies, the participants agreed on the following research questions building the basis for the analysis:

\begin{itemize}[noitemsep,topsep=0pt,partopsep=0pt]
\item \textbf{Q0,\brexitbold{}:} What can Wikipedia tell us about the UK's changing place in the world after Brexit?
\item \parisnew{\textbf{Q0,\parisbold{}:} Has the announcement of the US pullout of the Paris Agreement changed the depiction of and attention to the issue of climate change?}
\end{itemize}

\vspace{8pt}

In order to approach these research questions, the participants analyzed the following aspects in the course of their studies:

\begin{itemize}[noitemsep,topsep=0pt,partopsep=0pt]
\item \textbf{Q1:} How was the event-centric information propagated across languages?
\item \textbf{Q2:} How coherent are the articles regarding the event across languages?
\item \textbf{Q3:} Which aspects of the event are controversial across languages?
\end{itemize}

\subsection{Data Collection}
\label{sec:page_selection}

\begin{table}[t!]
\centering
\caption{Overview of the datasets resulting from the data collection.}
\label{tab:dataset}
\begin{tabular}{l|l|l||l|l|}
\cline{2-5} 
 & \bwdbold & \brefbold & \pabold & \pwdbold \\ \hline
\multicolumn{1}{|l|}{\textbf{\#Languages}} & $59$ & $48$ & $34$ & $4$ \\ \hline
\multicolumn{1}{|l|}{\textbf{\#Words (EN)}} & $4,468$ & $12,122$ & $5,950$ & $5,787$ \\ \hline
\multicolumn{1}{|l|}{\textbf{\#Categories (EN)}} &  $3$ & $9$ & $159$ & $9$ \\ \hline
\multicolumn{1}{|l|}{\textbf{\#Other Articles}} & \multicolumn{2}{c||}{-} & \multicolumn{2}{c|}{$99$} \\ \hline
\end{tabular}
\end{table}

First of all, in both case studies the participants selected a set of relevant articles to be analyzed, which resulted in the datasets shown in Table \ref{tab:dataset}.
In \brexit{}, the participants selected two English articles: the ``United Kingdom European Union membership referendum, 2016'' article (\bref{}) and the ``United Kingdom withdrawal from the European Union'' article (\bwd{}). 
For \paris{}, there is a ``Paris Agreement'' article (\pa{}) in several languages, but only four language editions provided a ``United States withdrawal from the Paris Agreement'' article (\pwd{}). Thus, the participants searched for paragraphs in Wikipedia articles linking to the articles \pa{} and ``Donald Trump'',  \pa{} and ``United States'', and those linking to \pwd{} using the Wikipedia API and manual annotation. To address \textbf{Q0,\parisbold{}}, the articles ``Global warming'', ``2015 United Nations Climate Change Conference'', ``Climate change'' and ``Climate change denial'' were considered.

\subsection{Analysis Methods and Feature Groups}
\label{sec:methods}

Overall, the analysis methods employed by the participants in both studies can be categorized into content, temporal, network and controversy analysis: 

\begin{itemize}[noitemsep,topsep=0pt,partopsep=0pt]
\item \textbf{Content analysis} to get detailed insights of how the event was described across languages (\textbf{Q0}).
\item \textbf{Temporal analysis} to analyze when sub-events were reported (\textbf{Q1}).
\item \textbf{Network analysis} to estimate the coherence between the event-centric articles across languages (\textbf{Q2}).
\item \textbf{Controversy analysis} to identify the controversial event aspects (\textbf{Q3}).
\end{itemize}

\vspace{8pt}

Table \ref{tab:analyses_and_feature_groups} provides an overview of the analysis methods and the corresponding features employed in both case studies. For clarity, we categorize the features into groups that were covered to a different extent in the case studies:

\begin{itemize}[noitemsep,topsep=0pt,partopsep=0pt]
\item \textbf{Text-based features:} Features based on the Wikipedia article texts such as the textual content, terms, selected paragraphs and the table of contents.
\item \textbf{Multimedia features:} Features based on the multimedia content (such as images) directly embedded in the articles.
\item \textbf{Edit-based features:} Features based on the editing process in Wikipedia, including the discussion pages and the editors.
\item \textbf{Link-based features:} Features employing the different types of links such as cross-lingual links between Wikipedia articles and links to external sources.
\item \textbf{Category-based features:} Features employing Wikipedia categories.
\end{itemize}

\input{table_methods_and_features}

\subsection{Observations}
\label{sec:feature_selection}

\subsubsection*{Content Analysis} 
Due to the language barrier, the participants of \brexit{} focused on the text-based features involving less text: tables of contents (TOCs) and images.
Similarly, in \paris{} the terms from the extracted paragraphs were utilized.

\begin{figure}[t!]
\floatbox[{\capbeside\thisfloatsetup{capbesideposition={right,top},capbesidewidth=0.43\textwidth}}]{figure}[\FBwidth]
{\caption{A comparison of the ToCs of the Brexit referendum article on the 24th June 2016 in four languages. The ToC entries are ordered by frequency and alphabetically. Darker colors correspond to a higher number of recurrences across languages, including standard Wikipedia sections and a section about the referendum results.}\label{fig:toc}}
{\includegraphics[width=0.55\textwidth]{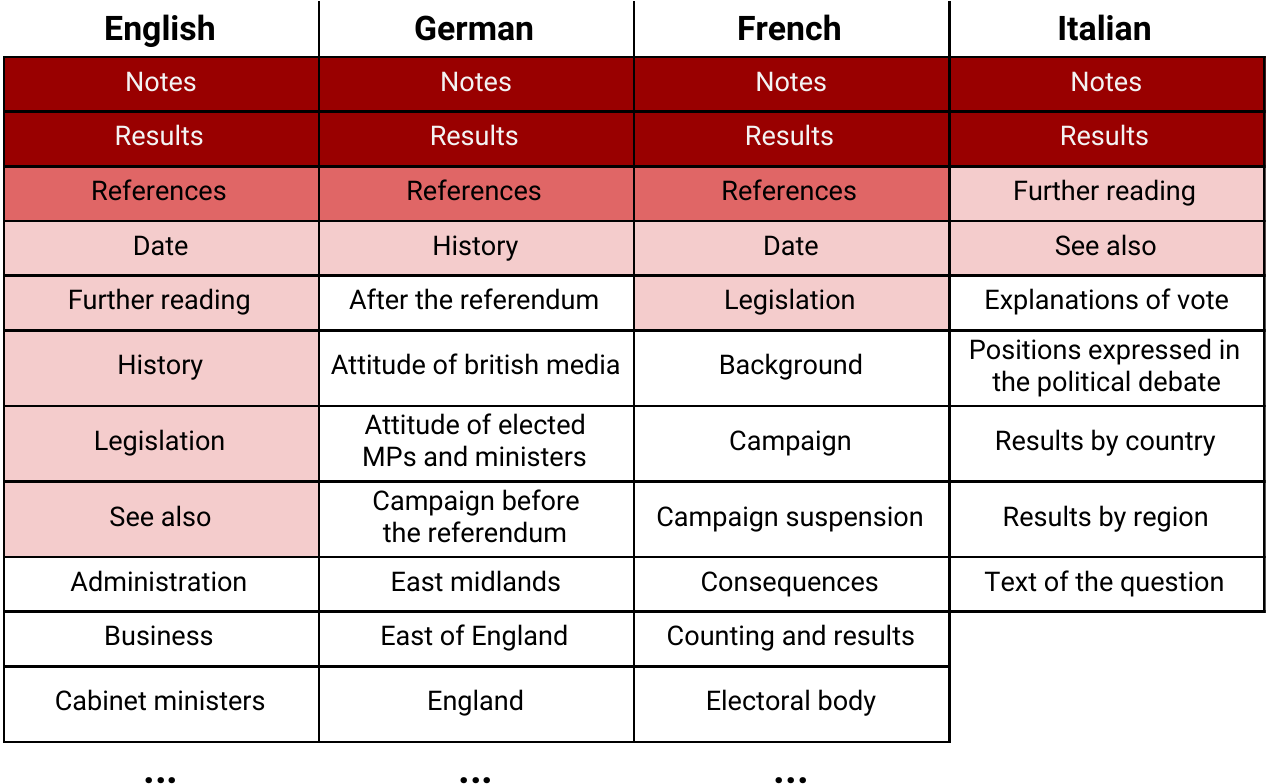}}
\end{figure}

\textit{Text-based features in \brexit{}}: 
The participants arranged the ToC entries by their frequency across languages as shown in Fig. \ref{fig:toc}. This ToC comparison indicates that the articles differ in many aspects, e.g. the German article focuses on the referendum results in different regions, and the English Wikipedia focuses on Brexit's economical and political implications. 

\textit{Multimedia features in \brexit{}}: Using the Wikipedia Cross-lingual Image Analysis tool\footnote{\url{wiki.digitalmethods.net/Dmi/ToolWikipediaCrosslingualImageAnalysis}}, it became evident that images containing the UK map and the referendum ballot paper were shared most frequently across languages.

\textit{Text-based features in \paris{}:} From the paragraphs extracted during the dataset collection, the participants extracted frequent words used in the context of the US pullout per language. This analysis showed different emphasis on the topic: The English Wikipedia mentioned oil and gas, the French Wikipedia included climate-related terms, and the Dutch one was focused on resistance.

\subsubsection*{Temporal Analysis} 
The description of ongoing events may vary substantially over time. 
In \brexit{}, the participants tracked this evolution using 
text-based and edit-based features. In \paris{}, the participants focused on multimedia features.

\textit{Text-based features in \brexit{}:} 
The participants extracted the ToC three times per day, in the time from the 22nd to the 24th of June 2016. The French version did not have the referendum results on the 23rd of June as opposed to other languages. On the following day, the English ToC stayed nearly unchanged, whereas a number of new German sections were added. 

\begin{figure*}[t!]
	\centering
	\includegraphics[width=\textwidth]{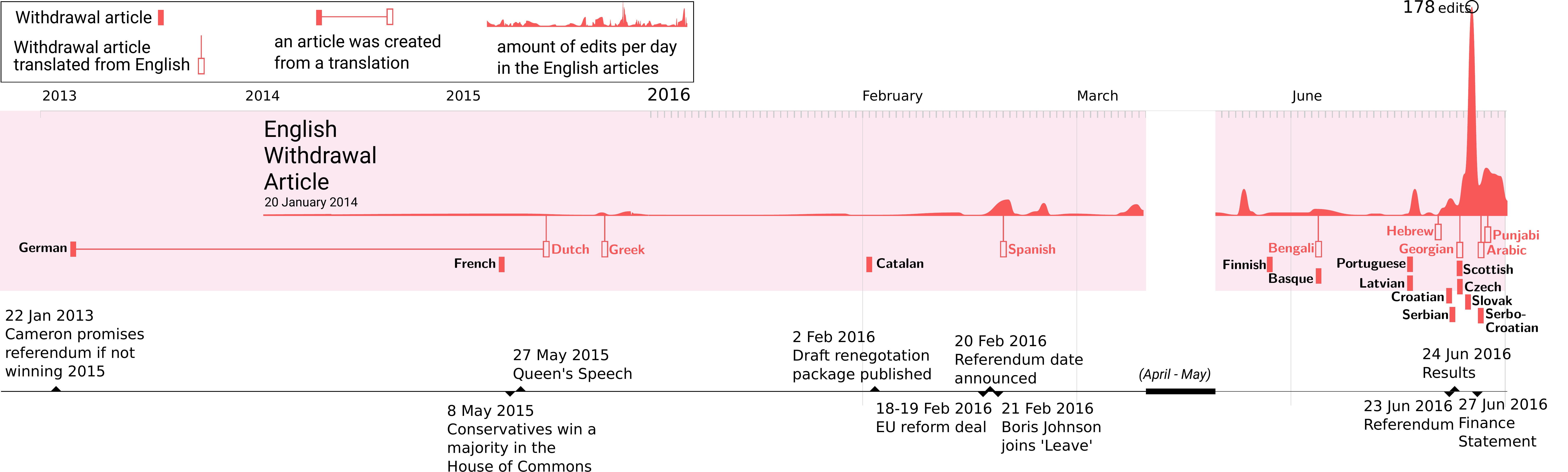}
	\caption{A timeline of the \bwd{} article showing the English edit frequency over time and article editions in other languages. For example, the Dutch article was created on the 16th of August 2015 as a translation from German and English.}
    \label{fig:timeline}
\end{figure*}

\textit{Edit-based features in \brexit{}} enabled observations of the Wikipedia editing process. The participants created a timeline depicted in Fig. \ref{fig:timeline} which is based on the data from the Wikipedia Edits Scraper and IP Localizer tool\footnote{\url{wiki.digitalmethods.net/Dmi/ToolWikipediaEditsScraperAndIPLocalizer}}. It illustrates the development of the \bwd{} article including its creation in other language editions. Article editions directly translated from other languages were marked and important events related to Brexit were added.

\textit{Multimedia features in \paris{}:} Motivated by \textbf{Q0,\parisbold{}}, the \paris{} participants compared the images added to the set of climate-related articles before and after the US pullout became apparent. The majority of newly added images reflect statistics (in contrast to a mixture of photos and statistics added before), and some of them depict the US influence on the world's climate.

\subsubsection*{Network Analysis} 
The coherence of the Wikipedia language editions can provide useful insights. The participants of \brexit{} focused on link-based and category-based features, while the participants of \paris{} focused on text-based features.

\begin{figure*}[t!]
	\centering
    \begin{subfigure}{0.49\textwidth}
   	\centering
	\includegraphics[width=1.0\textwidth]{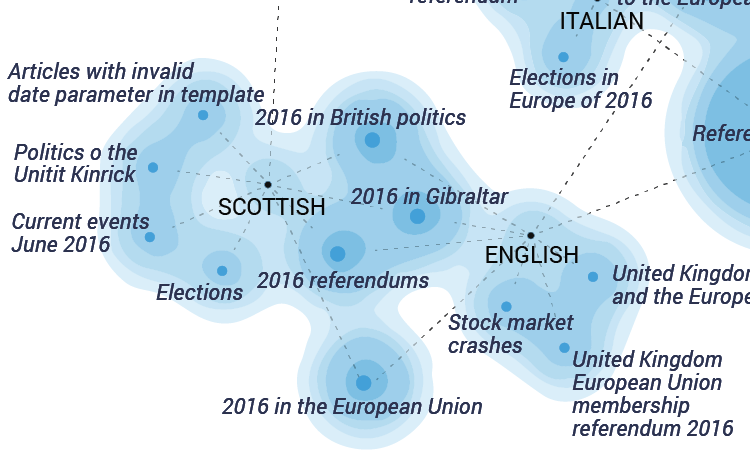}
    \caption{Categorization of the Brexit referendum article across languages with language nodes and Wikipedia category nodes. The edges represent connections between categories and languages.}
    \label{fig:categories_uk}
    \end{subfigure}\hfill
    \begin{subfigure}{0.49\textwidth}
    \centering
	\includegraphics[width=1.0\textwidth]{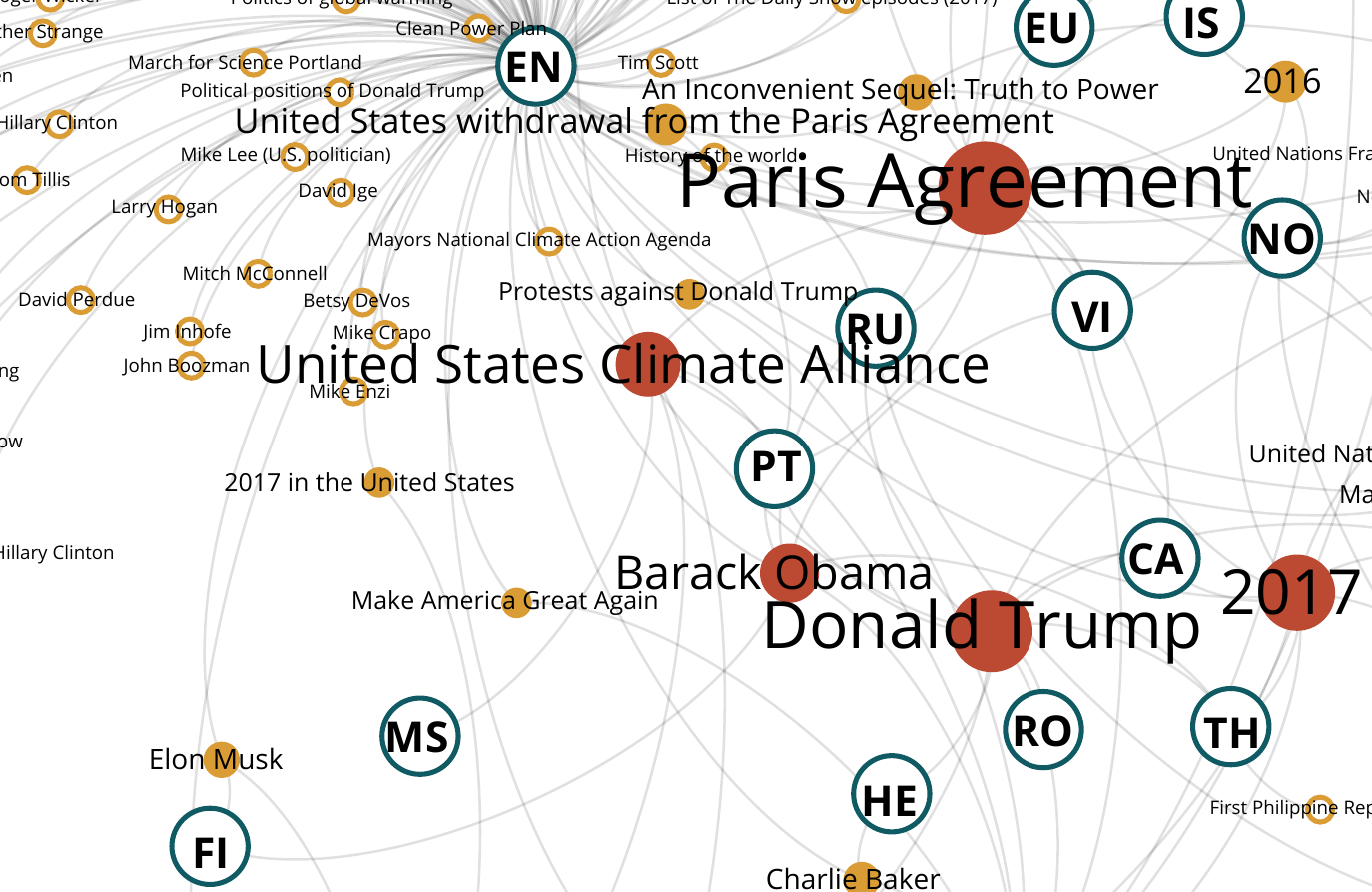}
    \caption{Articles mentioning the US pullout of the Paris Agreement. Blue nodes represent languages and the others Wikipedia articles, where color and size denotes how many language editions link to the article.}
    \label{fig:paris_network}
    \end{subfigure}\hfill 
    \caption{Network analysis in \brexit{} (category-based) and \paris{} (text-based).}
    \label{fig:network_analysis}
\end{figure*}

\textit{Category-based features in \brexit{}:}
The participants analyzed the categorization of the referendum and withdrawal articles in all available language editions by inserting the translated and aligned category names into the Table2Net tool\footnote{\url{http://tools.medialab.sciences-po.fr/table2net/}} and applying the ForceAtlas algorithm \cite{Jacomy:2014} to create the network shown in Fig. \ref{fig:categories_uk}, which shows an isolated position of the English and Scottish articles. 

\textit{Link-based features in \brexit{}:} Links to external sources were extracted and compared using the MultiWiki tool \cite{GottschalkD17}. For most of the language editions, the overlap of the linked web pages was rather low and reached higher values only in few cases, e.g. the English and German withdrawal articles shared $17.32\%$ of links.

\textit{Text-based features in \paris{}:} The set of articles mentioning the US pullout of the Paris Agreement was put in a network shown in Fig.~{\ref{fig:paris_network}}. ``Donald Trump'', ``Paris Agreement'', ``2017'' and ``United States Climate Alliance'' are the articles mentioning the withdrawal in most languages, while some articles such as ``Elon Musk'' only mention it in very few languages. Another observation was the separation of political and science-related articles.

\textit{Multimedia features in \paris{}:} \parisnew{The participants retrieved a list of images used in the different language versions of the ``Climate Change'' article. A network where language nodes were connected to images revealed that some language editions (e.g. Dutch) and groups of languages (e.g. a group of Serbian, Slovakian, Serbo-Croatian and Faroese) differed from the others with respect to the image use.}

\subsubsection*{Controversy Analysis:} In \brexit{}, the participants observed controversies among the Wikipedia editors. In \paris{}, no explicit controversy analysis was conducted due to the difficulties to resolve the origins of the extracted text paragraphs, the language barrier and the lack of extraction tools.

\textit{Edit-based features in \brexit{}:} For each Wikipedia article, there is a discussion page, structured by a table of contents. The \brexit{} participants reviewed the English discussion TOCs and identified an intense discussion among the Wikipedia editors on the question to which article the search term “Brexit” should link to.

%% file: table_methods_and_features.tex
\begin{table}[t!]
	\caption{For each of the four analysis methods, this table lists whether the participants employed features from the specified feature group in \brexit{} or \paris{}.}
	\label{tab:analyses_and_feature_groups}
	\centering
	\small
	\begin{tabular}{|l|c|c|c|c|c|}\hline
		\diagbox[width=\dimexpr 0.26\textwidth+2\tabcolsep\relax]{\textbf{Analysis}}{\textbf{Feat. Group}} & \multicolumn{1}{l|}{\textbf{\begin{tabular}[c]{@{}l@{}}Text-\\ based\end{tabular}}} & \multicolumn{1}{l|}{\textbf{\begin{tabular}[c]{@{}l@{}}Multi-\\ media\end{tabular}}} & \multicolumn{1}{l|}{\textbf{\begin{tabular}[c]{@{}l@{}}Edit-\\ based\end{tabular}}} & \multicolumn{1}{l|}{\textbf{\begin{tabular}[c]{@{}l@{}}Link-\\ based\end{tabular}}}  & \multicolumn{1}{l|}{\textbf{\begin{tabular}[c]{@{}l@{}}Cate-\\ gories\end{tabular}}} \\ \hline
		\textbf{Content} & \brexit{}, \paris{} & \brexit{} &  & &  \\ \hline
		\textbf{Temporal} & \brexit{} & \paris{} & \brexit{} & \no{}  & \\ \hline
		\textbf{Network} & \paris{} & \paris{} & \no{} & \brexit{} & \brexit{} \\ \hline
		\textbf{Controversy} & \no{} & \no{} & \brexit{} & \no{} & \no{} \\ \hline
	\end{tabular}
\end{table}

%% file: 06_discussion_new.tex
\section{Discussion}
\label{sec:discussion}

In this section we discuss our observations performed in the course of the case studies with respect to the objectives $O1$-$O3$ (Section \ref{sec:objectives}).

\subsection{\textit{O1-O2}: Influence Factors for Content, Methods and Feature Selection}

Overall, we observed that the adopted methods and their outcomes are influenced and also 
limited by a number of factors, including: 

\textbf{Characteristics of the event.} 
Relevant characteristics of the event include its topical breadth and  global influence. For example, as the Brexit referendum was considered as an event of European importance, the participants of \brexit{} focused on the European languages, while in \paris{} the US pullout of the Paris Agreement was studied in all languages due to its global impact. As the US pullout was considered to cover many aspects in politics and science, the participants focused on coverage, resulting in a larger set of articles to be analyzed compared to \brexit{}. 

\textbf{Participant professional background.}  
The professional background of the participants influenced in particular the selection of analysis methods. Although the teams were interdisciplinary, the individual participants 
focused on the methods and features typical for their disciplines. 

\textbf{Participant language knowledge.} 
The participant language knowledge limited in particular their ability to apply analysis methods that require close reading. For example, the controversy analysis in \brexit{} was only performed on the English discussion pages due to the language barrier. 
Nevertheless, the overall scope of the study was not limited by this factor. To cross the language barrier, the participants employed two techniques: 1) machine translation tools, and 2) content selection to reduce the amount of information in a foreign language to be analyzed (e.g. analyzing category names instead of full text). 

\textbf{Availability of the analysis tools.} 
Existing analysis tools 
mostly support distant reading on larger collections using specific features, such as links, images, etc. and can be applied in the multilingual settings efficiently. 
Fewer tools, including for example MultiWiki \cite{Gottschalk:2016}, support close reading in the cross-lingual settings. The edit-based and text-based cross-lingual controversy analysis was not adequately supported by the existing tools.

\textbf{Information structure.} 
One important factor of the event-centric cross-lingual analytics 
is the information structure. The features adopted in the case studies under consideration include rich text-based features such as hyperlinks and categories, as well as edit histories available in Wikipedia.
Furthermore, the availability of comparable articles in different languages is an important feature of the Wikipedia structure in this context.

\subsection{\textit{O3}: Findings, Limitations and Recommendations}


Table \ref{tab:findings_and_limitations} provides an overview of the analysis methods.  

\begin{table}[!t]
\scriptsize
\caption{Analysis methods: findings, limitations and tool recommendations.}
\label{tab:findings_and_limitations}
\begin{tabular}{|l|l|l|l|}
\hline
\textbf{Method} & \textbf{Facilitated findings} & \textbf{Current limitations} & \textbf{Recommendations} \\ \hline
Content & \begin{tabular}[c]{@{}l@{}}- Shared article aspects \\ - Interlingual dependencies\\ - Overview of the context \end{tabular} 
& \begin{tabular}[c]{@{}l@{}}- Lack of shared fact analysis\\ - Lack of systematic content \\ \ \ \  selection for close reading \end{tabular} 
& \begin{tabular}[c]{@{}l@{}}- Cross-lingual fact alignment\\ - Overview as an entry\\ \  point to close reading
\end{tabular}
\\ \hline
Temporal & \begin{tabular}[c]{@{}l@{}}- Event dynamics\\ - Changes in public interest\\ - Language version originality \\ - Context shift\end{tabular} &
\begin{tabular}[c]{@{}l@{}}- Analysis is limited to \\ \  specific revisions 
\end{tabular} & \begin{tabular}[c]{@{}l@{}} - Higher temporal resolution \end{tabular} \\ \hline
Network & \begin{tabular}[c]{@{}l@{}}- Cross-lingual similarity \\ - Event coverage \end{tabular} & \begin{tabular}[c]{@{}l@{}}- Roles of specific commu- \\ \ \ nities are hard to identify \end{tabular} & - Editor community insights  \\ \hline
\begin{tabular}[c]{@{}l@{}}Contro- \\ versy \end{tabular} & - Event perception & \begin{tabular}[c]{@{}l@{}}- Lack of cross-lingual\\ \ \ discussion comparisons \end{tabular} & - Discussion insights \\ \hline
\end{tabular}
\end{table}

\textbf{Findings:} The adopted analysis methods facilitated a range of findings. 
Content analysis using ToCs, images and word clouds enabled the identification of shared aspects, interlingual dependencies and provided a context overview. 
Temporal analysis involving ToCs, edit histories and images provided insights into the event dynamics, changes in the public interest within the language communities, originality of the language versions and the context shift. 
Network analysis resulted in an overview of the cross-lingual similarities, supported identification of the roles of the language editions, event coverage and the specific cross-lingual aspects. 
Controversy analysis conducted on the English Wikipedia (only) provided details on the event perception.

\textbf{Limitations:} 
The limitations of the adopted analysis methods mostly regard 
the relatively low content and temporal resolution, as well as the lack of detailed insights into the communities and discussions behind the content.
With respect to the content analysis, the lack of close reading restricted the obtained insights to rather high-level comparative observations, such as shared aspects of the articles, rather than individual facts.
In the temporal analysis, the information regarding content propagation was restricted to the origin of the first revision of the articles. 
The network analysis did 
not support insights in the specific communities behind the edits, such as the supporters and the opponents of Brexit.
The controversy analysis based on discussion pages could not be applied in the cross-lingual settings,
due to the lack of specific extraction tools and the language barrier.
Overall, the limitations observed in our study are due to multilingual information overload, the language barrier and the lack of tools to systematically extract and align meaningful items (e.g., facts) across languages.

\textbf{Recommendations:} Our observations regarding the above limitations and the post-sessions interviews lead to the recommendations for future method and tool development summarized in Table \ref{tab:findings_and_limitations}.
These recommendations include a zoom-out/in functionality to provide an overview, helping to select relevant content for close reading, extraction and cross-lingual alignment of information at a higher granularity level (e.g. facts), tracking article development over time including involved communities as well as a systematic analysis of discussion pages to better support controversy detection.
In the future work we would also like to develop interactive cross-lingual search and exploration methods based on \cite{DemidovaZN13}.

\subsection{Limitations of the Study}
In this qualitative study we limited our corpus to the multilingual Wikipedia, such that features and tools adopted by the participants are in some cases corpus-specific. 
Nevertheless, we believe that the results with respect to the participant strategies (e.g. the preferential usage of metadata to reduce close reading in a foreign language) are generalizable to other multilingual event-centric corpora, such as event-centric materials extracted from the Web corpora and Web archives.

%% file: 07_related_work.tex
\section{Related Work}
\label{sec:related}

Multilingual and temporal analytics becomes an increasingly important topic in the research community (see e.g. \cite{Kim:2016}, \cite{Govind:2017}
for recent studies on cross-lingual content propagation and editing activity of multilingual users).
Till now only few studies focus on analyzing and effectively supporting the needs of researchers who 
conduct research on temporal content \cite{Fernando2018a}, \cite{Odijk15} and 
create event-centric temporal collections \cite{GossenDR17}, \cite{GossenDR15}, \cite{risse2014collect}. Whereas existing works on temporal collections focus on the monolingual case, working patterns and requirements of researchers analyzing multilingual temporal context remain largely non-investigated. 

As Wikipedia language editions evolve independently and can thus reflect community-specific views and bias, multilingual Wikipedia became an important research target for different disciplines. One important area of interest in this context is the study of differences in the linguistic points of view in Wikipedia (e.g. \cite{Rogers:2013}, \cite{AlKhatib:2012}). Whereas several visual interfaces and interactive methods exist to support researchers in analysing Wikipedia articles across languages (e.g. MultiWiki \cite{GottschalkD17}, Contropedia \cite{Borra:2015b}, Manypedia \cite{Massa:2012} and Omnipedia \cite{Bao:2012}), 
our case study illustrates that substantial further developments are required to effectively support researchers in various aspects of cross-lingual event-centric analysis.

%% file: 08_conclusion.tex
\section{Conclusion}
\label{sec:conclusion}
In this paper we presented two case studies in which we observed interdisciplinary research teams who conducted research on the event-centric information in the context of the Brexit referendum and the US pullout of the Paris Agreement.
We summarized our observations regarding the 
content, method and feature selection, their influence factors as well as 
findings facilitated by the adopted methods and provided recommendations for services that can better support cross-lingual analytics of event-centric collections in the future.